\documentclass{article}

\usepackage{microtype}
\usepackage{graphicx}
\usepackage{subfigure}
\usepackage{booktabs}
\usepackage{hyperref}
\usepackage{subcaption}  
\usepackage{amsmath}
\usepackage{amssymb}
\usepackage{mathtools}
\usepackage{amsthm}
\usepackage[capitalize,noabbrev]{cleveref}
\usepackage[textsize=tiny]{todonotes}
\usepackage{booktabs}

\usepackage[accepted]{icml2025}

\theoremstyle{plain}

\theoremstyle{definition}

\theoremstyle{remark}

\icmltitlerunning{Generation of structure-guided pMHC-I libraries using Diffusion Models}

\begin{document}

\twocolumn[
\icmltitle{Generation of structure-guided pMHC-I libraries using Diffusion Models}

\begin{icmlauthorlist}
\icmlauthor{Sergio Emilio Mares}{ccb}
\icmlauthor{Ariel Espinoza Weinberger}{eecs}
\icmlauthor{Nilah M. Ioannidis}{ccb,eecs,biohub}
\end{icmlauthorlist}

\icmlaffiliation{ccb}{Center for Computational Biology, University of California, Berkeley, Berkeley, California, USA}
\icmlaffiliation{biohub}{Chan Zuckerberg Biohub, San Francisco, California, USA}
\icmlaffiliation{eecs}{Department of Electrical Engineering and Computer Sciences, University of California, Berkeley, Berkeley, California, USA}

\icmlcorrespondingauthor{Sergio Emilio Mares}{sergio.mares@berkeley.edu}
\icmlkeywords{Computational Biology, Generative AI, Peptide Design, MHC Class I, Immunotherapy, Machine Learning, ICML}

\vskip 0.3in
]

\printAffiliationsAndNotice{}

\begin{abstract}
Personalized vaccines and T-cell immunotherapies depend critically on identifying peptide-MHC class I (pMHC-I) interactions capable of eliciting potent immune responses. However, current benchmarks and models inherit biases present in mass-spectrometry and binding-assay datasets, limiting the discovery of novel peptide ligands. To address this issue, we introduce a structure-guided benchmark of pMHC-I peptides designed using diffusion models conditioned on crystal structure interaction distances. Spanning 27 high-priority HLA alleles, this benchmark is independent of previously characterized peptides yet reproduces canonical anchor residue preferences, indicating structural generalization without experimental MS dataset bias and reduced systematic bias. Using this resource, we demonstrate that state-of-the-art sequence-based predictors perform poorly at recognizing the binding potential of these structurally stable designs, indicating allele-specific limitations invisible in conventional evaluations. Our geometry-aware design pipeline yields peptides with high predicted structural integrity and higher residue diversity than existing datasets, representing a key resource for unbiased model training and evaluation. Our code and data are available at: \href{https://github.com/sermare/struct-mhc-dev}{https://github.com/sermare/struct-mhc-dev}.
\end{abstract}

\section{Introduction}
Peptide--MHC class~I (pMHC-I) interactions are central to adaptive immunity, enabling cytotoxic T~cells to recognize and eliminate infected or cancerous cells \cite{Chaplin2010, Hilf2019}. Predictors of pMHC-I binding have become widely used tools for personalized T-cell immunotherapies and modern vaccine design \cite{Saxena2025}. Given the vast combinatorial diversity of $>100,000,000$ distinct peptides, and after accounting for polymorphisms, insertions, deletions, and aberrant splicing, experimentally mapping all binding peptides is infeasible \cite{Yewdell2003_MakingSense}. Accurate algorithmic predictions are essential in vaccine trial design and necessitate precise \emph{in silico} prediction methods to prioritize candidate peptides for immunotherapeutic development \cite{Hilf2019,Walz2015}. Despite substantial progress, current pMHC-I prediction methods face important limitations. Most state-of-the-art models are sequence-based and trained on a large library of known binders from public databases, such as the Immune-Epitope Database (IEDB) which contains a library of $>10^6$ pMHCs \cite{Vita2025}. These datasets predominantly originate from mass-spectrometry (MS) immunopeptidomics \cite{Sarkizova2020} and \emph{in~vitro} binding assays and thus carry experimental biases. One well-documented bias is the under-detection of cysteine-containing peptides in standard MS workflows, which in turn causes such peptides to be under-represented in databases and often missed by trained predictors \cite{Bruno2023, dincer2022reducing, mallick2007computational}. Furthermore, many benchmarks rely on similar experimental data for evaluation, potentially inflating performance by testing on peptide sequences with distributions similar to model training sets. This over-reliance on biased datasets and narrow test sets raises concerns that reported accuracy overstates real-world generalization capacity \cite{Machaca2024}.

To address these challenges, we introduce a structure-conditioned diffusion model for pMHC-I peptide generation. Our diffusion-based generative model explicitly conditions on the three-dimensional structure of the MHC-I binding groove. By leveraging this structural context, it designs peptides predicted to be compatible with the binding pocket of a given MHC allele, increasing their predicted structural validity. This approach enables exploration of peptide sequence space beyond the biases present in current databases, yielding novel out-of-distribution peptides guided by structural binding preferences. By generating plausible yet unconventional peptides, our model expands the landscape of candidate peptide sequences and provides challenging new test cases for evaluating existing predictors.

\begin{figure*}[ht]
\vskip 0.2in
\begin{center}
\includegraphics[width=0.9\textwidth, trim=10 160 10 210, clip]{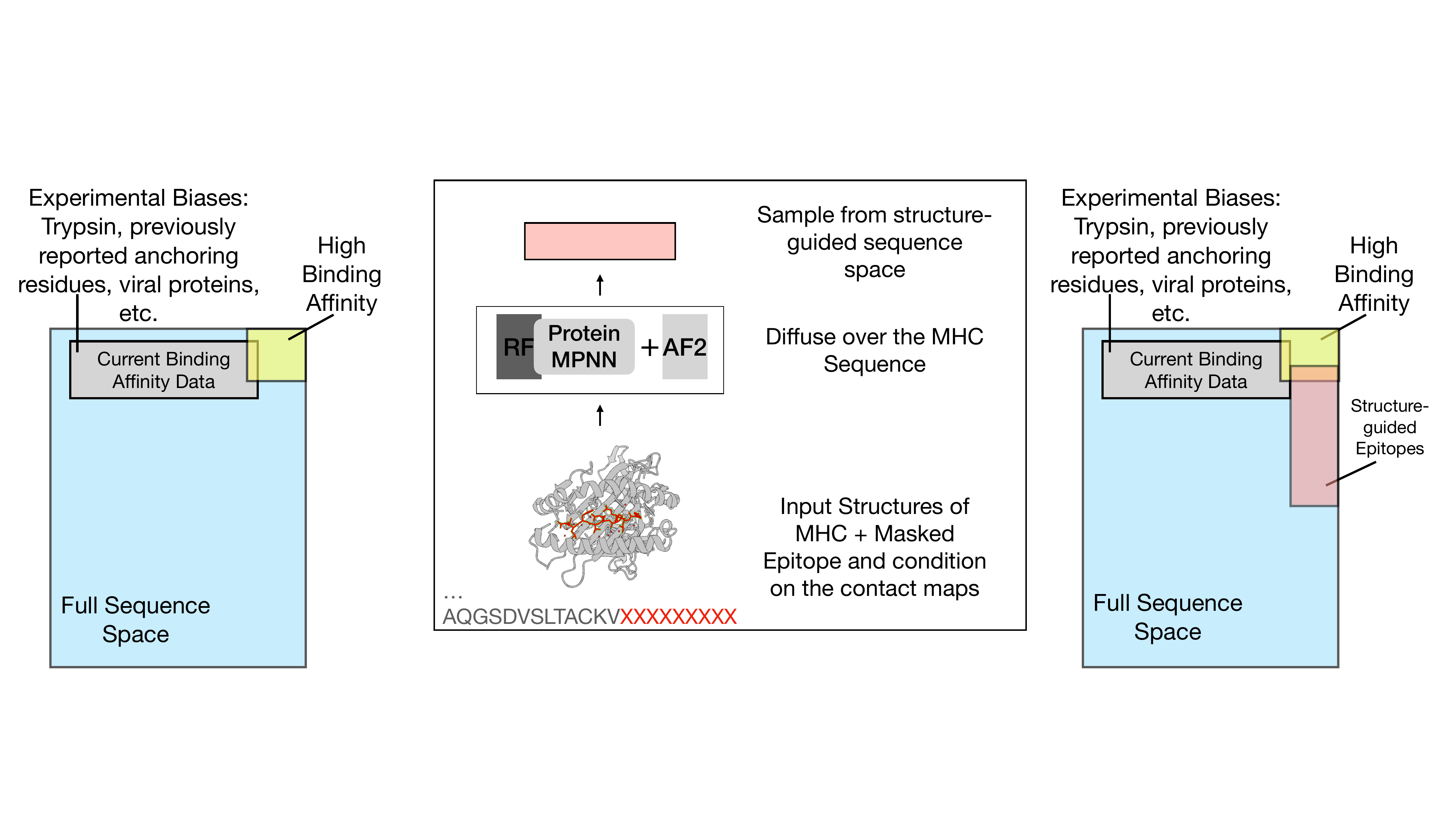}
\caption{Overview of the structure-guided generative pipeline for designing high-affinity peptides for MHC class I molecules.}
\label{fig:pipeline}
\end{center}
\vskip -0.2in
\end{figure*}
\vspace{0.3em}

\section{Data and Methods}

\subsection{Dataset}

We collected 189 peptide--MHC class~I crystal structures from the Protein Data Bank (PDB) \cite{Berman2003}, covering 27 distinct human HLA alleles, accessed 23~April~2025. Only peptides of length 9--11 amino acids were included. Structures were required to have crystallographic resolution $\leq 3.5$~\AA, providing sufficient detail to resolve hydrogen bonds and van der Waals contacts typically observed at this distance threshold. We excluded structures containing missing peptide residues, non-standard amino acids, incomplete HLA annotations, or redundant complexes of identical peptide--allele pairs with a final set of 119 pMHC structures (Supplementary Table~\ref{table:MHCPDBs}).

\paragraph{Contact computation.}
For each peptide residue $i$ and each MHC residue $j$, we calculated the minimum heavy-atom distance across all atoms of residues $i$ and $j$, recording this value as $d_{ij}^{\min}$. We systematically evaluated thresholds from 1.0 to 10.5~\AA\ in 0.5~\AA\ increments to characterize the distribution of pMHC proximities. While distances greater than 6--7~\AA\ rarely represent direct physical interactions, including the full range allowed us to observe the decay of pairwise contacts with distance.  

We followed a 3.5~\AA\ threshold to define hydrogen bonds and direct atom--atom contacts observed in pMHC structures \cite{Chaurasia2021, Anjanappa2020, Li2016}, while slightly larger cutoffs up to 5.0~\AA\ have been applied to capture looser pMHC associations \cite{sydney2016PNAS}. Approximately 5.7\% of pMHC residue pairs fall within $\leq 3.5$~\AA\ and 16.8\% within $\leq 5.0$~\AA. Based on these empirical distributions and established interaction criteria, we classified residue pairs into two categories: \emph{close contacts} when $d_{ij}^{\min} \leq 3.5$~\AA, typically corresponding to van der Waals interactions and candidate hydrogen bonds, and \emph{extended contacts} when $3.5 < d_{ij}^{\min} \leq 5.0$~\AA, representing weaker, less direct associations at the binding interface. 

\paragraph{Validation of distance cutoffs.}
To assess the robustness of these thresholds, we examined the empirical distribution of minimum heavy-atom distances across the dataset and confirmed stability of the observed proportions through bootstrap resampling. The 3.5~\AA\ cutoff consistently captured the closest and most structurally constrained residue pairs, whereas the 5.0~\AA\ cutoff provided broader coverage of the binding interface. Contact count distributions further indicated that individual peptide residues typically formed 1--4 close contacts ($\leq 3.5$~\AA) with MHC residues.    

\subsection{Generative Pipeline}

Our pipeline (Fig.~\ref{fig:pipeline}) begins with the crystallized MHC--peptide structure. We first remove the peptide coordinates, preserving the MHC scaffold and positional definitions of hot-spot residues. RFdiffusion \cite{watson2023} was run for 50 diffusion steps to generate peptide backbone candidates, with hot-spot contacts enforced as geometric constraints to encourage high-affinity binding. We specified a peptide length window of 9--11 residues, did not impose symmetry, and used a sampling temperature of 0.5. 

Each RFdiffusion-generated backbone was next threaded through ProteinMPNN \cite{dauparas2022} to assign peptide side-chain identities while keeping the MHC scaffold fixed. The peptide coordinates were isolated, while the entire complex backbone was provided to ProteinMPNN in \verb|--complex| mode. For each backbone, we sampled $N=64$ sequences at a temperature of 0.5, with cysteine residues disabled (\verb|rm_aa=C|) to prevent disulfide-driven artifacts. Sequences were ranked by their negative log-likelihood (NLL), and the top five were retained for structural evaluation.

The top ProteinMPNN sequences were folded with AlphaFold2-Multimer (AF2) \texttt{binder} with the HLA heavy chain and $\beta$\textsubscript{2}-microglobulin kept intact. AF2 was run with \texttt{model\_1\_multimer\_v3}, \texttt{num\_recycles}=6, and without templates. For each design, we recorded the mean per-residue pLDDT over peptide positions and the interface pTM (iPTM) score. Peptides with pLDDT $\geq 0.80$ were retained as high-confidence designs. 

\subsection{Diffusion model refinement and bias}
The authors of RFdiffusion and ProteinMPNN have previously reported amino-acid composition biases in generative design tasks, such as a strong preference for proline in loop scaffolds where Rosetta-based methods failed \cite{dauparas2022}. Such behavior, while sometimes beneficial for rigidifying flexible regions, can also reflect unintended model artifacts. Similar biases may arise in peptide design tasks, leading to overrepresentation of certain residues across alleles or positions. To evaluate potential model-driven signals in our structure-guided pMHC peptides, we quantified amino-acid usage both globally and at each sequence position for each HLA allele. Enrichment was then assessed relative to a position-specific background estimated from the high-confidence anchoring set (Supp. Fig.~\ref{fig:biasinrf}).

\subsection{Position Weight Matrix Normalization}
To account for amino acid usage biases, we normalized the enrichment plots using background frequencies from the human proteome. For the sequence logo plots, the position-specific probability matrices were corrected by dividing by the corresponding background frequencies, ensuring that observed enrichments reflect deviations from both the proteome baseline and biases introduced by the diffusion model.

\subsection{Models Evaluated}
 Binding affinities are predicted by MHC-Flurry \cite{odonnell2020}, NetMHCpan \cite{mcinnes2018}, HLApollo \cite{Thrift2024}, HLAthena \cite{Sarkizova2020}, MixMHCpred \cite{BassaniSternberg2017_DecipheringHLAI}, MHCNuggets \cite{Shao2020_MHCnuggets},  and ESMCBA, a fine-tuned ESM-Cambrian model \cite{mares2025continueddomainspecificpretrainingprotein}.

\subsection{Additional Benchmarking Datasets}
We employed three additional benchmarking  datasets: 1. IEDB database with peptides after 2020, eliminating data leakage for most models' training data. 2. A constructed dataset with preservation of the anchoring residues following the same distribution of peptides observed in the public dataset, and the randomly generated rest of the sequence. 3. A list of 9-11-mers auto-regressively generated with ESM2 \cite{ESM2}, starting from an initial random token and sampling each residue from the model's predictive distributions. We ensured independence of the dataset by removing any overlapping peptides between this generated pipeline and the public databases.

\section{Results}

\subsection{Allele-aware peptide similarity}

To investigate whether generated peptides recapitulate allele-specific binding preferences, we constructed Position Weight Matrices (PWMs) from peptides with high-confidence structures. Pairwise Jensen–Shannon (JS) distances were then computed between allele-specific PWMs to quantify motif similarity. 

The resulting distance matrix (Supp. Fig.~\ref{fig:SHANNONDIVERGENCE}) reveals that peptides generated for the same allele cluster together, with lower intra-allele divergences compared to inter-allele comparisons. Broad allele family structure is  visible; HLA-A and HLA-B, and within-family similarities (e.g., A02 alleles) are stronger than across families (e.g., HLA-A vs HLA-B).

\begin{figure}[ht]
  \vskip 0.1in
  \begin{center}
    \includegraphics[width=0.47\textwidth, trim=0 0 0 0 , clip]{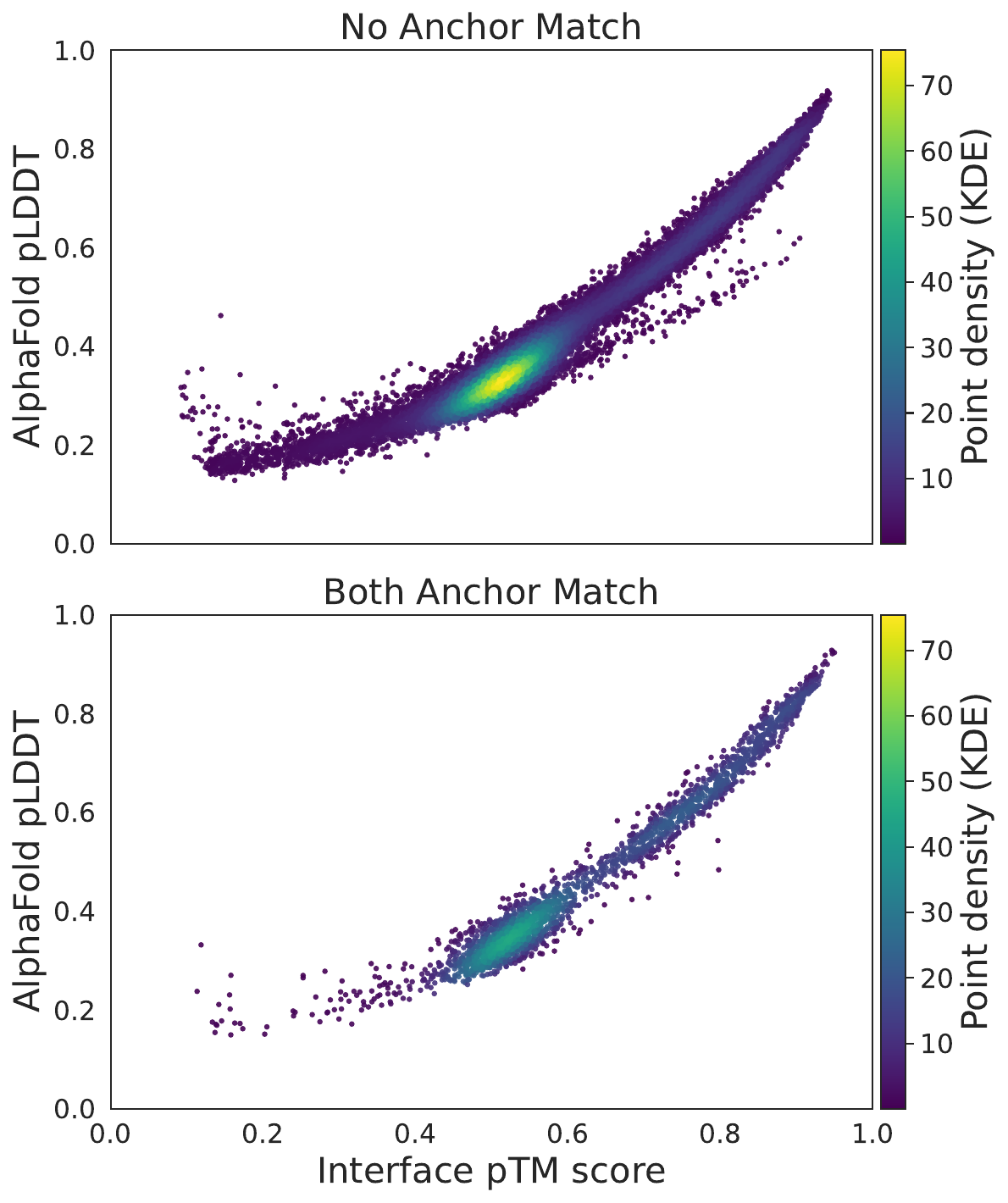}
    \caption{Predicted interface pTM scores and AlphaFold pLDDT values for peptides with and without anchor matches.}
    \label{fig:iptm_scatter}
  \end{center}
  \vskip -0.1in
\end{figure}

\subsection{Recapitulating canonical anchors align with higher predicted structural confidence}

We tested whether generated peptides containing canonical anchor residues yielded higher-confidence structural predictions. Canonical anchors are expected to recover the characteristic MHC-I binding motif and, if correctly captured, should also yield more confident model predictions compared to peptides without anchors. 

The interface predicted TM-score (iPTM) is a metric from AlphaFold that estimates the accuracy of inter-chain packing in protein–protein or protein–ligand complexes, providing a confidence measure for modeled interfaces. Alongside per-residue pLDDT, we used iPTM to assess model certainty. The presence of canonical anchors systematically shifted the distribution toward higher iPTM and pLDDT scores (Fig.~\ref{fig:iptm_scatter}). We evaluated HLA-A*02:01 peptides lacking anchors displayed broader, lower-confidence distributions, whereas those containing both position-2 (P2) and $P\Omega$ anchors were enriched with higher iPTM and pLDDT, with a highly significant difference confirmed by the Mann--Whitney test ($p < 4.5 \times 10^{-75}$).

These results demonstrate that canonical anchors not only recover expected MHC-I binding motifs but also enhance model confidence in structure predictions.

\subsection{Experimental Validation with orthogonal unbiased data}

EpiScan is a high-throughput, cell-based platform that presents bar-coded peptide libraries on the surface of MHC-I molecules and quantifies their relative presentation by deep sequencing. Because it bypasses mass-spectrometry and in vitro binding assays, EpiScan provides an unbiased measurement of peptide presentation \cite{Bruno2023}. EpiScan's study tested $>$500,000 peptides and found $>$40,000 peptides that bound to HLA-A*02:01 and $>$17,000 for HLA-B*57:01. Our structure-guided diffusion library complements EpiScan by generating anchor-compatible peptides that explore under-sampled regions in silico, mitigating diversity limitations imposed by experimental sampling.

\paragraph{Enrichment in HLA-A*02:01 and HLA-B*57:01} %
In HLA-A*02:01, canonical aliphatic anchors are observed in EpiScan, with Valine strongly enriched at $P\Omega$ and Leucine enriched at both P2 and $P\Omega$ (Supp. Fig.~\ref{fig:a0201_episcan}). Our library recapitulates these canonical signals, recovering the Valine enrichment at $P\Omega$ and the Leucine signal at P2 (Supp. Fig.~\ref{fig:a0201_this_paper}).

In HLA-B*57:01, Tryptophan, a bulky aromatic hydrophobic residue, is the dominant canonical anchor at $P\Omega$. This enrichment is captured clearly in both EpiScan and our generated library, though with a reduced magnitude in the latter (Supp. Fig.~\ref{fig:b5701_episcan}). Threonine, a small polar residue, is also enriched at P2 and P8 in EpiScan. Although not enriched at the canonical classical P2 in our epitopes of this allele, we recover the P8 signal (Supp. Fig.~\ref{fig:b5701_this_paper}).

In addition to recovering canonical anchor motifs, we also observe allele-independent amino-acid biases that are consistent with previously reported artifacts in RFdiffusion and ProteinMPNN. These biases appear as low-variance enrichments across alleles (e.g., underrepresentation of cysteine and overrepresentation of certain loop-rigidifying residues), suggesting a model-driven rather than biological origin. To account for non-biological amino-acid biases introduced by the generative models, we corrected each allele-specific PWMs against a positional background distribution estimated from the full anchoring set. The resulting sequence logos highlight allele-specific enrichments above background rather than global design artifacts (Supp. Fig. \ref{fig:seq_logo}).

\subsection{Evaluating Novelty and Binding Potential Against Public Datasets}

\begin{figure}[ht]
    \vskip 0.2in
    \begin{center}
        \includegraphics[width=0.5\textwidth]{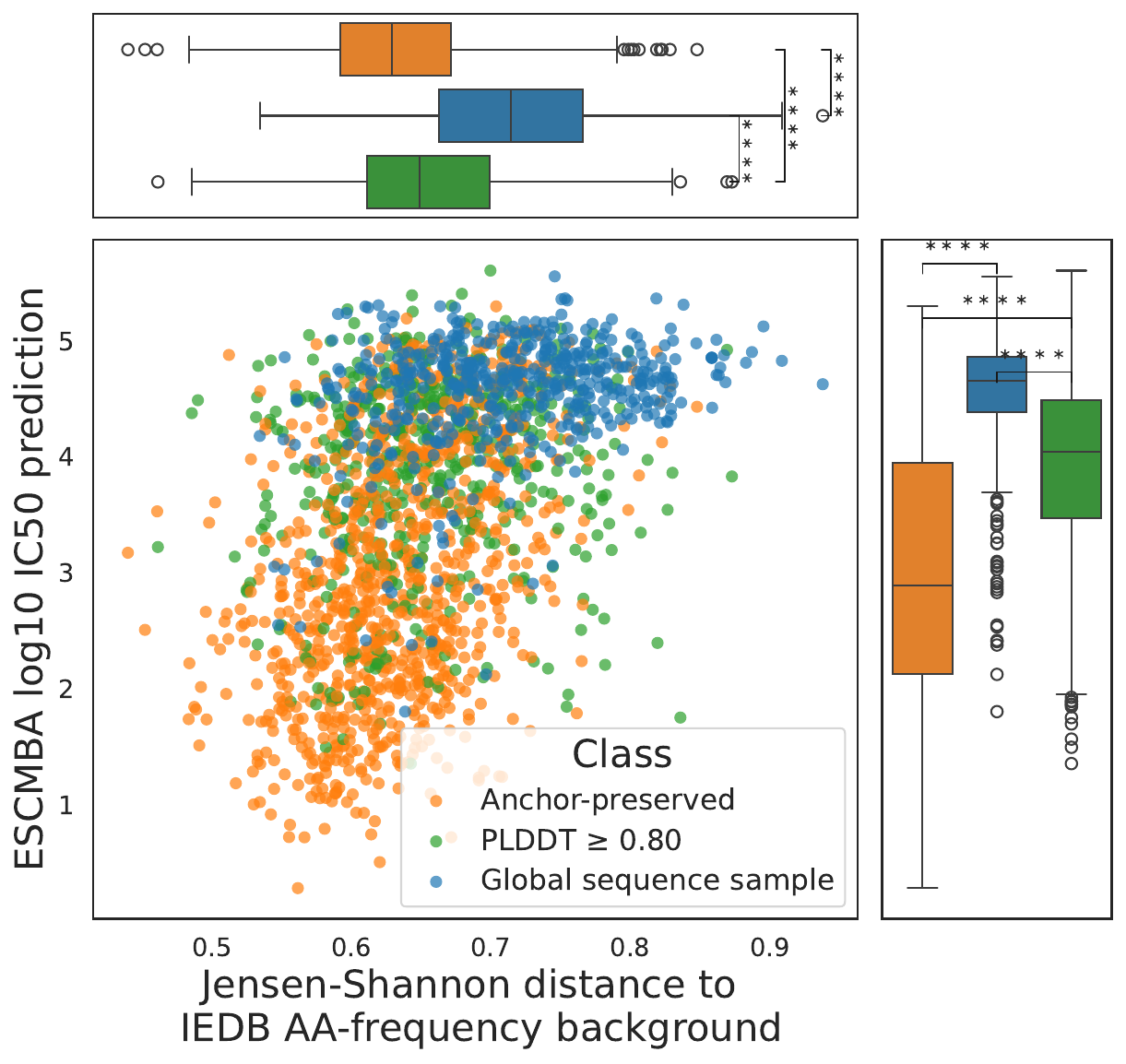}
        \caption{JS distance to the IEDB background and binding affinity predictions.}
        \label{fig:novelty_affinity}
    \end{center}
    \vskip -0.2in
\end{figure}

To quantify how our designed HLA-A*02:01 peptides differ from existing public epitopes, we compared their amino acid composition to the IEDB background using Jensen-Shannon distance while simultaneously evaluating predicted binding affinity. Anchor-preserved sequences are statistically closer to IEDB distributions and consistently display higher predicted binding (Fig.~\ref{fig:novelty_affinity}), whereas globally sampled and structure-guided sequences explore more novel composition space. These orthogonal metrics demonstrate that our approach balances novelty with functional potential, highlighting how structural constraints (anchor preservation) trade off against diversity and affinity, while unconstrained and confidence-filtered designs can escape known biases yet remain structurally plausible binders. 

Additionally, these generated peptides lie outside the training distribution, and thus their true binding capacity remains unknown. This is precisely the motivation of this study, which seeks to test whether such novel out-of-distribution sequences can function as binders.

\subsection*{Sequence Novelty and Latent Space of Generated Designs}

To further contextualize these compositional differences, we examined the latent embedding space of peptide sequences from the model’s final hidden layer using UMAP (Supplementary Fig.~\ref{fig:umap_emb}). This analysis supports the PWM-based distances: diffusion-generated epitopes cluster into distinct regions of latent space, separated from random controls and anchor-permutation baselines. Together, the PWM comparisons and embedding analyses demonstrate that our diffusion-based pipeline explores novel yet structured sequence space, extending beyond biases of current public datasets and traditional generative baselines.

\begin{figure*}[ht]
  \vskip 0.2in
  \begin{center}
    \includegraphics[width=0.8\textwidth, trim=0 0 0 0, clip]{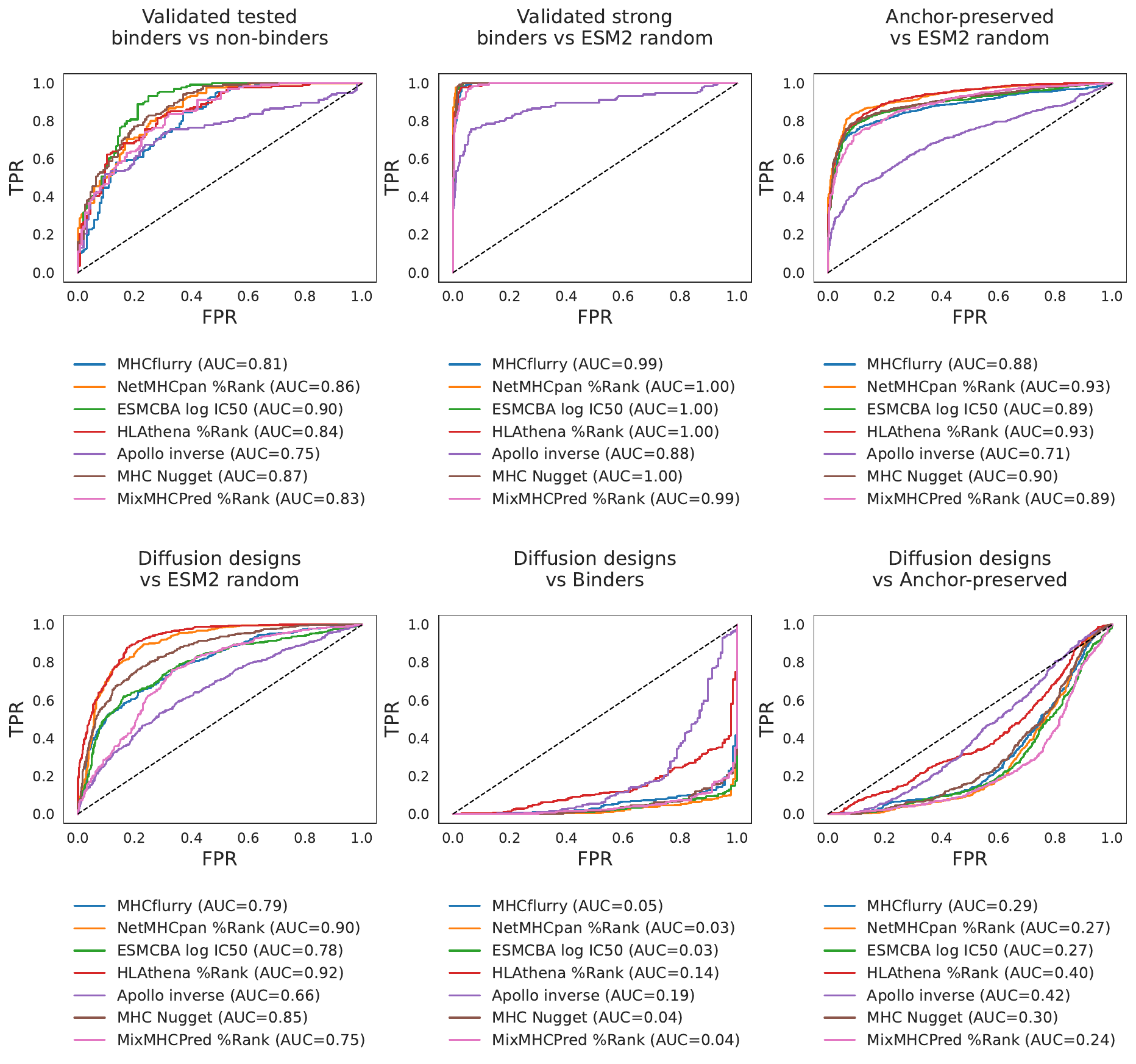}
    \caption{
ROC curves and AUROC values for seven binding affinity predictors across four peptide evaluation classes: experimentally validated binders, permutation-test controls, random-sampled negatives, and structure-guided diffusion designs.
}
    \label{fig:roc_curves}
  \end{center}
  \vskip -0.2in
\end{figure*}

\subsection{Full ROC curve performance across peptide classes} 

For each predictor, we constructed Receiver Operating Characteristic (ROC) curves across four distinct peptide evaluation classes and computed the Area Under the ROC (AUROC) to assess discriminative performance in HLA-A*02:01, since across alleles has the most training data in IEDB (Fig.~\ref{fig:roc_curves}). 

\subsection*{Performance on Experimentally Validated Binders}

To establish a baseline, we evaluated each predictor’s ability to distinguish experimentally validated binders from other peptides. All methods performed moderately, achieving AUROCs from 0.75 (Apollo) to 0.90 (ESMCBA), with most predictors clustering tightly around 0.74–0.90. These high AUROCs demonstrate robust recognition of known canonical binding motifs by current sequence-based methods, aligning well with their training data.

\subsection*{Performance on Random-Sampled Peptides}

To evaluate predictor specificity, we tested their ability to discriminate randomly sampled peptides from experimentally validated peptides with strong binding. Most predictors correctly assigned low binding scores to random peptides, achieving AUROCs from 0.88–1.00. These results confirm that predictors are robustly specific, effectively distinguishing random peptides from biologically plausible sequences.

\subsection*{Performance on Permutation-Test Peptides}

To test predictors' robustness to subtle sequence perturbations, we next assessed performance on anchor-preserved permutation-test peptides. Ideally, predictors should down-rank these controls, yielding AUROCs well above 0.5), reflecting their non-physiological sequence context. However, we observed AUROCs ranging from 0.71 (HLApollo) to 0.93 (HLAthena and NetMHCpan), indicating that predictors naively assigned these peptides relatively high scores. This result reveals a troubling sensitivity to subtle global sequence context perturbations beyond anchor positions, highlighting a critical vulnerability in current predictive approaches.

\subsection*{Performance on Structure-Guided Diffusion Designs}

Finally, we evaluated model performance on peptides explicitly designed to structurally complement the MHC binding pocket (structure-guided designs with pLDDT $>$ 0.8). We first evaluated them against the random generated peptides. All methods' AUROCs show clear ability to distinguish the diffusion peptides from random noise. However, all methods achieved poor discriminative performance against validated strong binders, with AUROCs ranging from 0.03 (MHC Nugget) to 0.19 (Apollo). This performance clearly exposes a significant blind spot: current predictors are largely unable to recognize structurally plausible peptides, highlighting critical limitations in their generalization capabilities and underscoring the need for structurally aware training data.

\section{Discussion and limitations}

Our study addresses critical limitations in pMHC-I binding prediction by leveraging diffusion models conditioned on atomic-level interactions, effectively avoiding biases inherent in mass-spectrometry and binding assay data. The structure-aware generative method introduced here challenges current models, highlighting their inability to recognize structurally valid, novel peptides that are out-of-distribution for their experimental training datasets. In particular, our approach recapitulates canonical anchoring preferences observed in unbiased datasets such as EpiScan (for example, recovery of tryptophan, valine, and leucine anchors) and shows that these residues systematically align with higher predicted structural confidence (iPTM and pLDDT). This anchor recovery provides confidence that our approach captures allele-specific motifs despite not being trained directly on binding data. At the same time, our methodology expands the explored sequence space, generating diverse peptides that probe regions of the epitope landscape underrepresented in current datasets.

Several limitations should be acknowledged. First, we observed a non-trivial enrichment of proline across alleles. This likely reflects the well-documented tendency of diffusion-based generative models to rigidify local structures by favoring proline, a behavior also noted by the authors of RFdiffusion in other contexts. While proline enrichment may provide stabilizing biophysical effects in some cases, its consistent appearance across unrelated alleles suggests that it can also represent a technical artifact of the generative process rather than a true biological signal. Second, our positional variance analysis revealed additional biases in amino-acid usage that were relatively uniform across alleles. Such across-allele uniformity may indicate model-driven design constraints that override allele-specific preferences, highlighting the importance of correcting for background biases when interpreting sequence logos or enrichment statistics. Third, although our approach successfully recapitulates canonical anchors and these residues were associated with systematically higher structural confidence, structural proxies remain imperfect surrogates for physical binding. The iPTM score, which measures AlphaFold2’s confidence in interface packing, and the per-residue pLDDT metric are both useful correlates, but they cannot directly prove biochemical stability or immunogenicity. Finally, by design our approach expands into sequence space that lies outside the distribution of known epitopes. This novelty is a strength, as it enables exploration of under-sampled regions of the binding landscape, but it also means that our generated peptides must ultimately be validated experimentally. Without functional testing, their true binding capacity and immunological relevance remain unknown.

Overall, our workflow and generated library provide both a benchmark for predictive models and a tool to uncover data- and model-driven biases. While our validation relied primarily on the unbiased EpiScan dataset, which is limited to four alleles, our framework is generalizable to a broader range of HLA types. Looking forward, incorporating TCR-binding predictions and extending experimental validation across additional alleles will broaden the relevance of this methodology for comprehensive immunotherapy design. In doing so, our approach establishes a path toward expanding the epitope landscape beyond current data limitations while maintaining biological plausibility through structural constraints.

\section{Data and Code Availability}

Data and code for all the analyses can be found in the Github repository: \href{https://github.com/sermare/struct-mhc-dev}{https://github.com/sermare/struct-mhc-dev}.

\section{Acknowledgements}
We thank Dr. Peter Bruno for valuable comments and help with the EpiScan analysis.

\bibliography{references}
\bibliographystyle{icml2025}

\appendix
\onecolumn

\section{Supplemental Tables}
\begin{table}[ht]
\centering
\begin{tabular}{|c|c|p{11cm}|}
\hline
\textbf{HLA Allele} & \textbf{\# Structures} & \textbf{PDB IDs} \\
\hline
A0201 & 15 & 7KGP, 7KGQ, 7LG2, 7MJ9, 7MKB, 7MJ6, 3TO2, 7KGO, 7KGS, 7MJ8, 7LG3, 7MJ7, 7KGT, 7SA2, 3OXS \\
\hline
A0206 & 1 & 3OXR \\
\hline
A0207 & 1 & 3OXS \\
\hline
A2    & 6 & 6UJQ, 6O4Z, 6O4Y, 6UJO, 6O51, 6O53 \\
\hline
A2402 & 7 & 7JYW, 7EJN, 8SBK, 7EJL, 7JYV, 7EJM, 8HN4 \\
\hline
A6    & 1 & 4HX1 \\
\hline
B1301 & 1 & 7YG3 \\
\hline
B1501 & 5 & 6UZS, 7XF3, 6VB3, 6UZQ, 6UZP \\
\hline
B1502 & 11 & 6VB6, 6VIU, 6VB7, 6UZM, 6UZO, 6VB1, 6UZN, 6VB0, 6VB4, 6VB2, 6VB5 \\
\hline
B1801 & 4 & 8ROP, 8RNH, 8ROO, 8RNG \\
\hline
B2705 & 7 & 6Y28, 6VQE, 6VQD, 6VQ2, 6VPZ, 6Y2A, 6Y26 \\
\hline
B2706 & 1 & 5DEG \\
\hline
B3501 & 7 & 8EMK, 8EMF, 7SIG, 8EMG, 8EMI, 8EMJ, 4PRA \\
\hline
B3503 & 1 & 7SIH \\
\hline
B3505 & 1 & 7SIF \\
\hline
B3508 & 13 & 4PRD, 3VFP, 3VFR, 3VFN, 3VFO, 3VFU, 7KGO, 3VFS, 3VFM, 3VFT, 4PRE, 3VFW, 3VFV, 4PRB \\
\hline
B4001 & 1 & 6IEX \\
\hline
B4403 & 2 & 3KPN, 3KPO \\
\hline
B4405 & 2 & 6MTL, 1SYV \\
\hline
B5301 & 2 & 7R7W, 7R7V \\
\hline
B5701 & 3 & 6BXQ, 7R7Y, 6V2O \\
\hline
B5703 & 2 & 5VVP, 5VWD \\
\hline
B5801 & 2 & 7X00, 7WZZ \\
\hline
B5803 & 1 & 5VWF \\
\hline
B8    & 3 & 8E8I, 8EC5, 8E13 \\
\hline
C1202 & 1 & 9L48 \\
\hline
C1203 & 1 & 9L49 \\
\hline
N0101 & 1 & 6K7T \\
\hline
\end{tabular}
\caption{HLA alleles with associated PDB structures}
\label{table:MHCPDBs}
\end{table}

\section{Supplemental Figures}

\renewcommand{\thefigure}{S\arabic{figure}}  
\setcounter{figure}{0}                      

\begin{figure}[ht]
\vskip 0.2in
\begin{center}
\includegraphics[width=1\textwidth, trim=0 0 0 0 , clip]{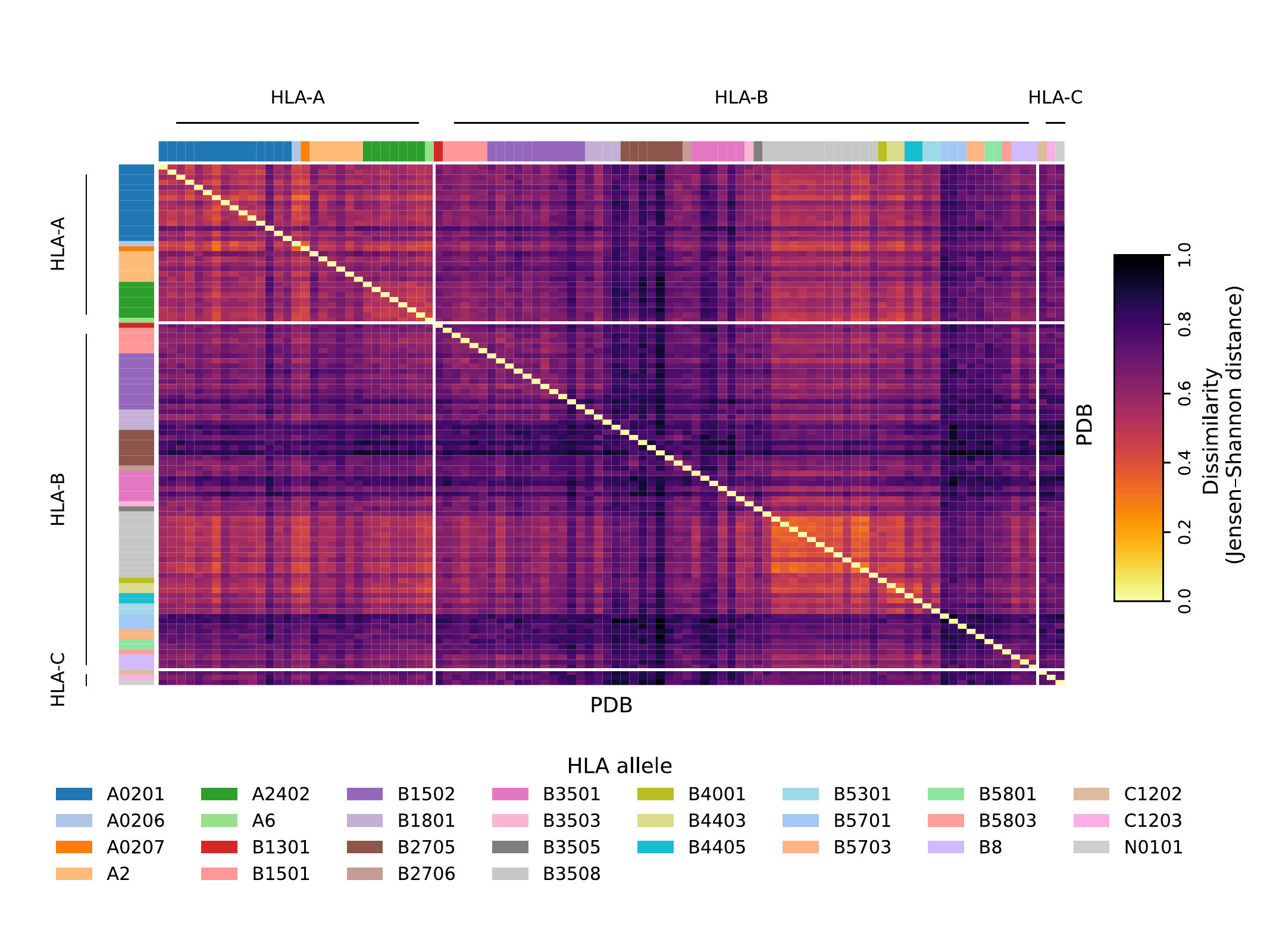}
\caption{Jensen-Shannon distance matrix between 9-mer PWMs extracted from high-confidence (pLDDT $\geq$ 0.7) from the structure-guided generated peptides.}
\label{fig:SHANNONDIVERGENCE}
\end{center}
\end{figure}

\begin{figure}[ht]
\vskip 0.2in
\begin{center}
\includegraphics[width=0.7\textwidth, trim=0 0 0 0 , clip]{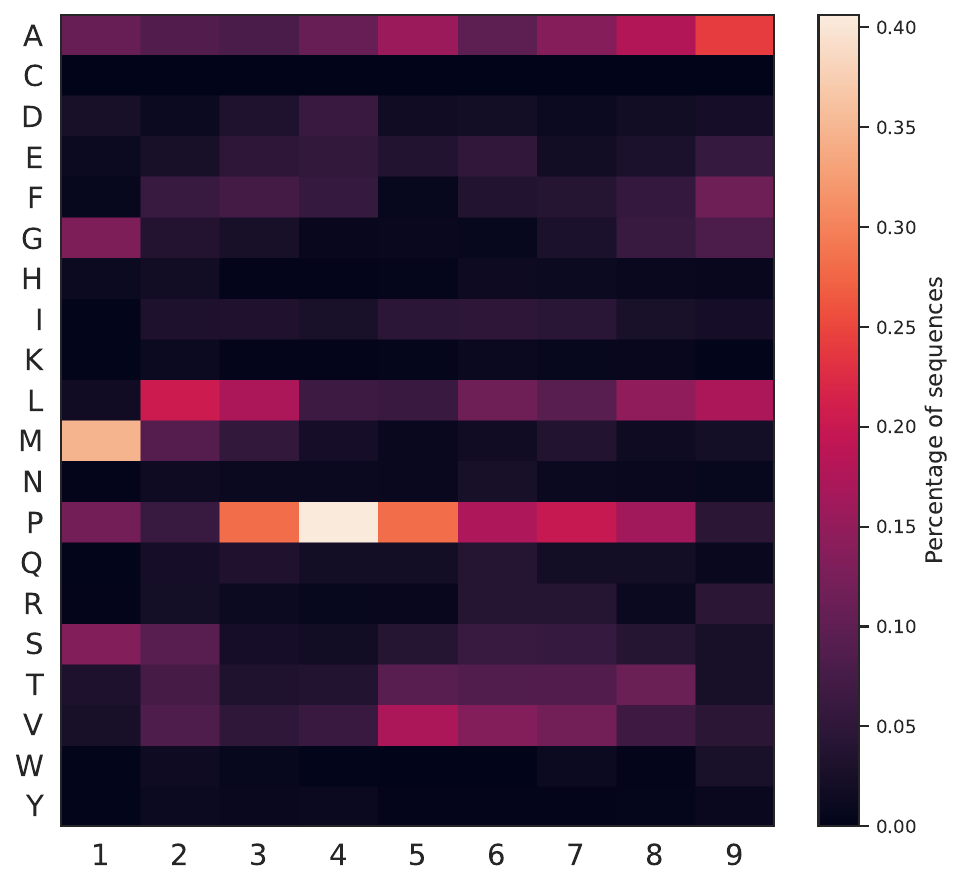}
\caption{Amino acid frequencies across positions in 9-mer structure-guided peptides from 27 alleles.}
\label{fig:biasinrf}
\end{center}
\end{figure}

\begin{figure}[ht]
  \centering
  \begin{subfigure}
    \centering
    \includegraphics[width=\textwidth]{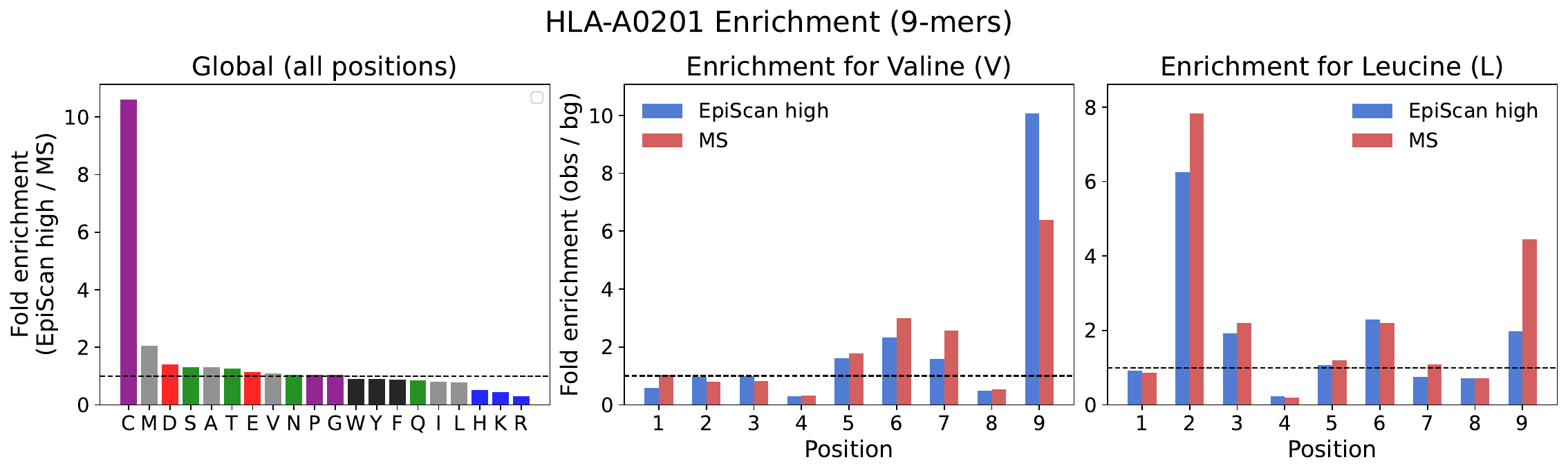}
    \caption{EpiScan results for HLA-A*02:01.}
    \label{fig:a0201_episcan}
  \end{subfigure}
  \hfill
  \begin{subfigure}
    \centering
    \includegraphics[width=\textwidth]{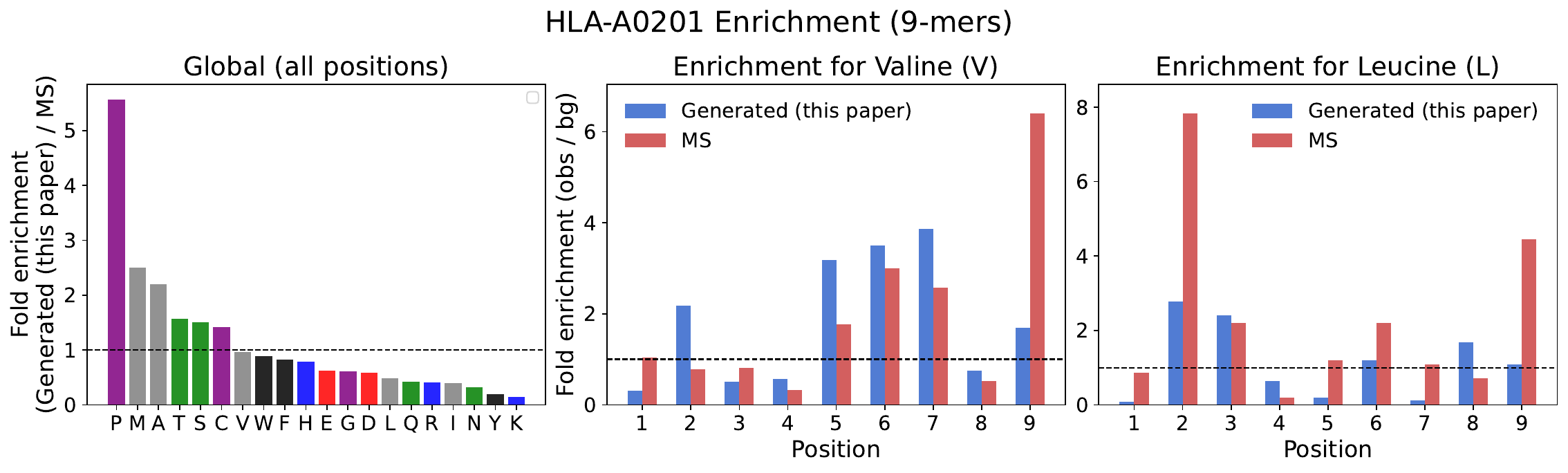}
    \caption{Results from this study for HLA-A*02:01.}
    \label{fig:a0201_this_paper}
  \end{subfigure}
  \label{fig:a0201_results}
\end{figure}

\begin{figure}[ht]
  \centering
  \begin{subfigure}
    \centering
    \includegraphics[width=\textwidth]{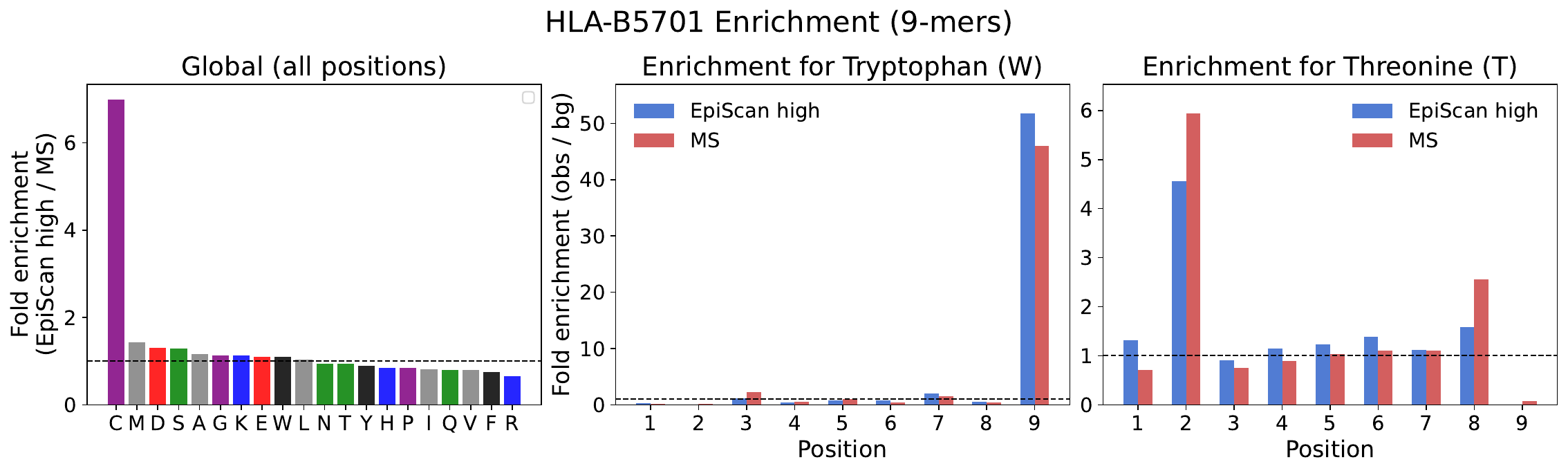}
    \caption{EpiScan results for HLA-B*57:01.}
    \label{fig:b5701_episcan}
  \end{subfigure}
  \hfill
  \begin{subfigure}
    \centering
    \includegraphics[width=\textwidth]{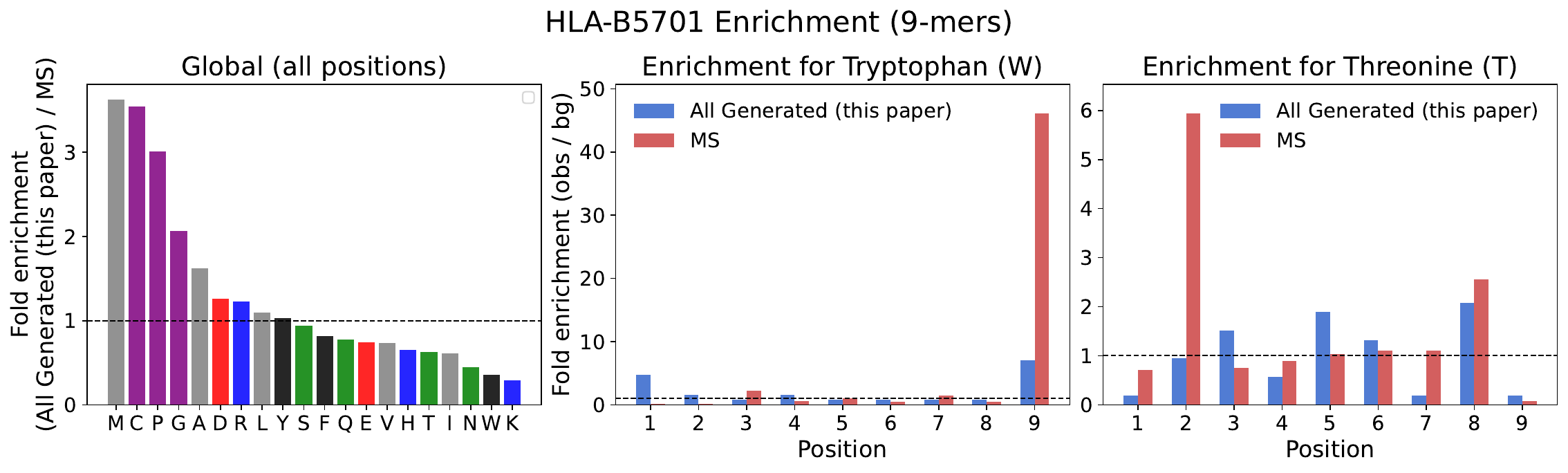}
    \caption{Results from this study for HLA-B*57:01.}
    \label{fig:b5701_this_paper}
  \end{subfigure}
  \label{fig:b5701_results}
\end{figure}

\begin{figure}[p]                 
  \centering
  \includegraphics[width=1\textwidth]{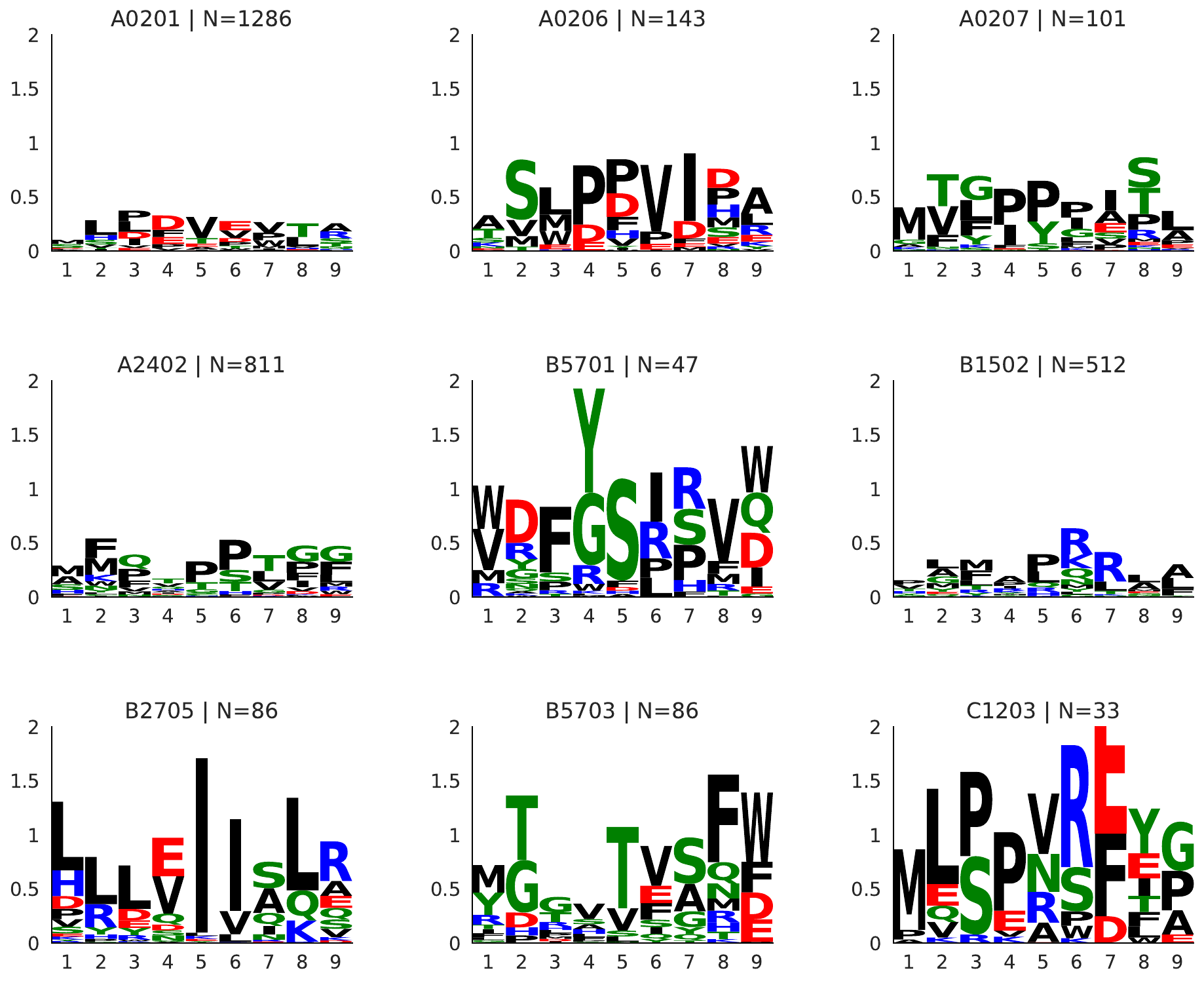}
\caption{Sequence logos of 9-mer peptides grouped by allele, normalized against positional background frequencies. The height of each letter reflects relative information content.}
  \label{fig:seq_logo}           
\end{figure}

\begin{figure}[ht]
\vskip 0.2in
\begin{center}
\includegraphics[width=1\textwidth]{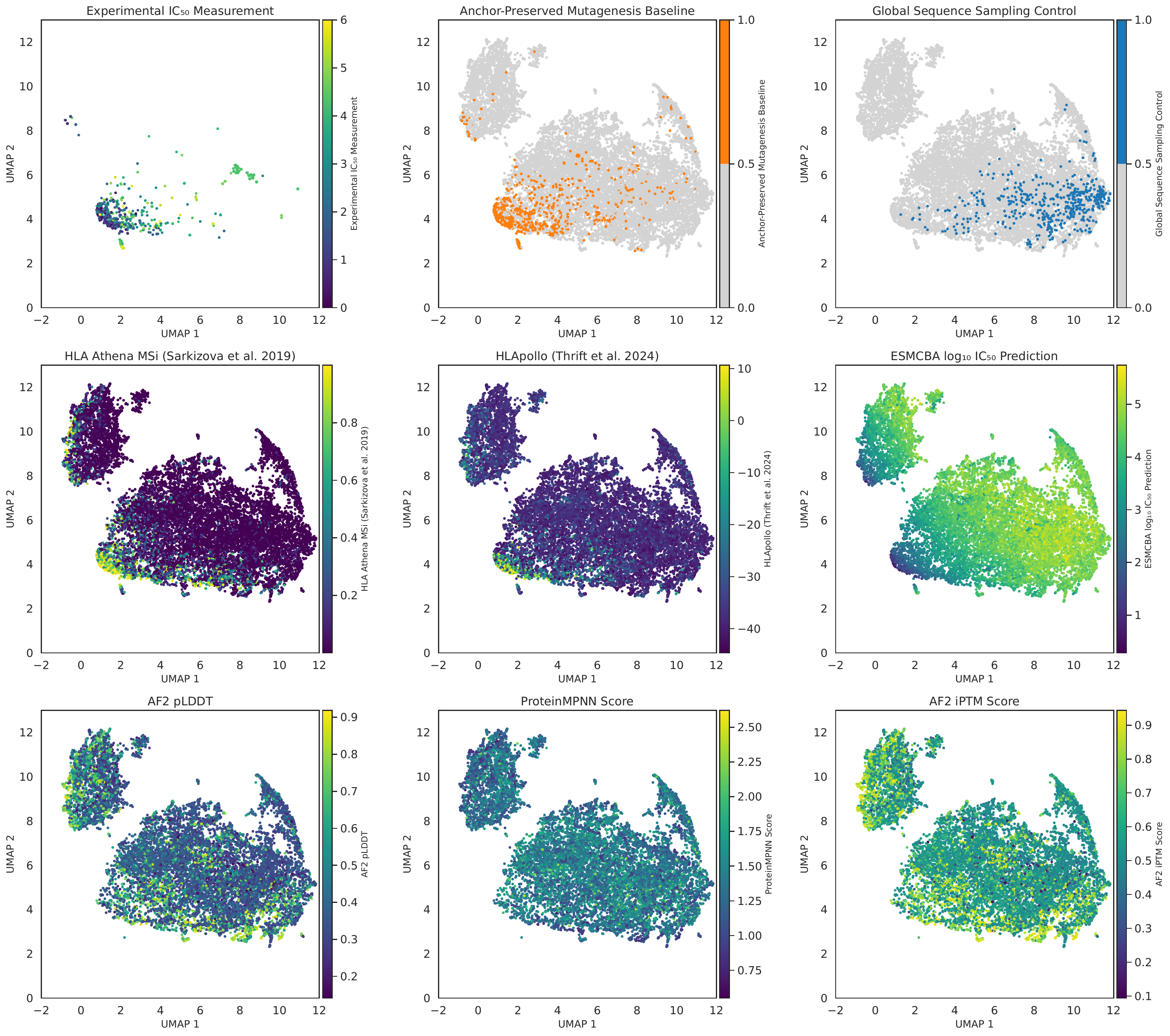}
\caption{UMAP visualization of ESMCBA embeddings for HLA-A*02:01 peptides across classes and scoring metrics. The first row shows the key peptide categories for sequence space analysis: Experimental IC$_{50}$ Measurement (experimentally validated epitopes), Anchor-Preserved Mutagenesis Baseline (control peptides with conserved anchor residues but randomized non-anchor positions), and Global Sequence Sampling Control (random peptides generated by autoregressive ESM2 sampling).}
\label{fig:umap_emb}
\end{center}
\vskip -0.2in
\end{figure}

\end{document}